\documentclass{article}
\usepackage{spconf,amsmath,graphicx,hyperref}
\usepackage{amssymb}

\title{An Overview of Rate-Distortion-Perception Theory}
\name{Jun Chen$^\dagger$, Ashish Khisti$^\ddag$}
\address{$^\dagger$ McMaster University, $^\ddag$ University of Toronto}
\begin{document}
%
\maketitle
\begin{abstract}

This paper provides a comprehensive overview of the rate-distortion-perception  framework, tracing its evolution from mathematical theory to practical deployment. We review the mathematical definition of the the mathematical definition of the  Blau--Michaeli  function and the associated coding theorems, examine computational methods for its evaluation, and discuss alternative formulations of the framework. Operational principles and design guidelines for modern image and video coding are also presented, highlighting the rate-distortion-perception framework as a rigorous foundation for developing next-generation perceptual compression systems.

\end{abstract}
\section{Introduction}
\label{sec:intro}

Traditional lossy compression is  rooted in Shannon’s rate-distortion theory, which characterizes 
the minimum achievable reconstruction distortion for a given rate budget.
 This framework has provided the conceptual and mathematical foundation for decades of image and video compression. However, distortion-based criteria such as mean-squared error do not fully capture human perceptual quality. In particular, at low bitrates, optimizing distortion alone often leads to overly smooth, blurry, or perceptually unnatural reconstructions. The rate-distortion-perception   framework addresses this limitation by introducing perceptual quality as a third dimension of the fundamental compression tradeoff. Rather than characterizing compression solely through rate and distortion, the new framework explicitly accounts for the statistical and perceptual fidelity of the reconstruction distribution. This seemingly natural extension leads to a substantially richer theory, in which the  performance limit depends delicately on how the perception constraint is formulated and imposed. Moreover, unlike classical rate-distortion theory, common randomness can play an essential role under certain perception constraints. These distinctions highlight the fundamentally different nature of perceptual compression, which goes beyond a simple refinement of classical distortion-based compression and opens up a distinct line of information-theoretic investigation.

Rate-distortion-perception theory lies at the intersection of information theory, optimal transport, and modern machine learning. On the theoretical side, this framework gives rise to new  coding theorems that characterize the performance limits of perceptual compression. Its connection to optimal transport and generative modeling
 offers a natural perspective on how perceptually realistic reconstructions can be produced under stringent rate constraints. On the computational side, evaluating these performance limits presents its own challenges and motivates a range of algorithms and optimization techniques. At the system level, the theory provides guidance for designing perceptual objectives, architectures, and coding strategies for modern image and video compression.
In this paper, we provide an overview of the rate-distortion-perception framework, covering its theoretical foundations, coding theorems, computational methods, and alternative formulations that reveal the richness of the framework and the subtleties of perceptual constraints. We further discuss how the theory provides a principled basis for leveraging emerging generative and learned compression techniques to develop next-generation perceptual compression systems.







\section{Blau--Michaeli Function}
\label{sec:format}

\subsection{Definition}

Shannon's distortion-rate function
\begin{align} D(R):=&\inf\limits_{p_{\hat{X}|X}}\mathbb{E}[d(X,\hat{X})]\\ &\mbox{s.t.}\quad I(X;\hat{X})\leq R
\end{align}
characterizes the minimum achievable reconstruction distortion under a prescribed rate budget $R$.
Here, $d:\mathcal{X}\times\hat{\mathcal{X}}\rightarrow\mathbb{R}_+$ is a distortion measure that quantifies the difference between the source and reconstruction symbols.

However, it has long been recognized in image and video coding that distortion-based measures do not necessarily align with human perception. Developing robust, non-reference measures of perceptual quality has been  a longstanding challenge, with many traditional attempts achieving only limited success. The recent rapid development of generative modeling, from generative adversarial networks and diffusion models to flow matching, offers a powerful new perspective on this problem. 
Since generative models aim to capture the underlying distribution of natural images, perceptual quality can be characterized in terms of the statistical distributions of the source and reconstruction. Specifically, by viewing the source and reconstruction domains as being governed by two probability distributions, $p_X$ and $p_{\hat{X}}$, respectively, perceptual quality assessment then boils down to measuring the discrepancy between these distributions through a divergence function $\phi$.

Against this backdrop, Blau and Michaeli \cite{BM19} introduced 
\begin{align} D(R,P):=&\inf\limits_{p_{\hat{X}|X}}\mathbb{E}[d(X,\hat{X})]\label{eq:objective}\\ &\mbox{s.t.}\quad I(X;\hat{X})\leq R,\\ &\hspace{0.3in}\phi(p_X,p_{\hat{X}})\leq P,\label{eq:single_perception}
\end{align}
which augments Shannon's distortion-rate formulation with an explicit perception constraint. This formulation intends to capture the fundamental tradeoff among compression rate, reconstruction distortion, and perceptual fidelity. In particular, it makes explicit the tension between faithful reconstruction of individual source realizations and statistical consistency with the source distribution, thereby providing an information-theoretic framework for analyzing perceptual compression.

\subsection{Coding Theorems}

At its inception, the operational meaning of the Blau--Michaeli function was not entirely clear. This was subsequently clarified through the work of Theis and Wagner \cite{TW21} and Chen et al. \cite{CYWSGT22}.

Let the source $\{X_t\}_{t=1}^{\infty}$ be an i.i.d. process with marginal distribution $p_X$. A length-$n$ perceptual compression system consists of a stochastic encoder $f^{(n)}:\mathcal{X}^n\times\mathcal{K}\rightarrow\mathcal{J}$, a stochastic decoder $g^{(n)}:\mathcal{J}\times\mathcal{K}\rightarrow\hat{\mathcal{X}}^n$, and a shared random seed $K$, uniformly distributed over $\mathcal{K}$ and independent of the source. The stochastic encoder $f^{(n)}$ maps the source sequence $X^n$ and the random seed $K$ to a codeword $J\in\mathcal{J}$ according to a conditional distribution $p_{J|X^nK}$, while the stochastic decoder $g^{(n)}$ generates the reconstruction $\hat{X}^n$ from $J$ and $K$ according to a conditional distribution $p_{\hat{X}^n|JK}$. A distortion level $D$ is said to be achievable under compression rate constraint $R$, perception constraint $P$, and common randomness rate constraint $C$ if there exists a sequence of length-$n$ perceptual compression systems satisfying: 
	\begin{align}
    &\frac{1}{n}\sum\limits_{t=1}^n\mathbb{E}[d(X_t,\hat{X}_t)]\leq D,\label{eq:distortion}\\
		&\frac{1}{n}\log|\mathcal{J}|\leq R,\label{eq:rate}\\
		&\frac{1}{n}\sum\limits_{t=1}^n\phi(p_{X},p_{\hat{X}_t})\leq P,\label{eq:perception}\\
        &\frac{1}{n}\log|\mathcal{K}|\leq C.\label{eq:commonrandomness}
	\end{align}
The infimum of all achievable distortions $D$ is denoted by $D_{\mathrm{M}}(R,P,C)$. Here, the subscript $\mathrm{M}$ indicates that perceptual quality is evaluated in terms of the marginal distributions of the source and reconstruction symbols, distinguishing this formulation from alternative perception criteria discussed later.


It was shown\footnote{All coding theorems stated in this paper are established under certain regularity conditions, which are typically mild. For brevity, we do not state these conditions explicitly; readers are referred to the original references for the precise assumptions underlying each result.} in \cite[Theorem 3]{TW21}, by leveraging the strong functional representation lemma \cite{LEG18}, that $D_{\mathrm{M}}(R,P,C)$ coincides with the Blau--Michaeli  function when an unlimited amount of common randomness is available, i.e.,
\begin{align}
D_{\mathrm{M}}(R,P,\infty)=D(R,P). \label{eq:infiniteC}
\end{align}
Subsequently, it was shown in \cite[Theorems 2,3,4]{CYWSGT22} that common randomness is actually unnecessary:
\begin{align}
D_{\mathrm{M}}(R,P,0)=D(R,P).\label{eq:zeroC}
\end{align}
In fact, as shown in \cite[Theorem 2]{CYWSGT22}, when perfect perception is not required (i.e., $P>0$), both the encoder and decoder can be chosen to be completely deterministic. The proof of \eqref{eq:zeroC} in \cite{CYWSGT22} relies on a delicate construction of codebooks with cyclic symmetry and a coding scheme that exploits this structure.

The proof in \cite{CYWSGT22} further indicates that \eqref{eq:zeroC} continues to hold when the perception constraint \eqref{eq:perception} is replaced by the more stringent per-symbol constraint
\begin{align}
\phi(p_X,p_{\hat{X}_t})\leq P,\quad t=1,2,\ldots,n.
\end{align}
Moreover, \eqref{eq:zeroC} remains valid under the following two variants of \eqref{eq:perception}:
\begin{align}
\phi\left(p_X,\frac{1}{n}\sum\limits_{t=1}^np_{\hat{X}_t}\right)\leq P
\end{align}
and
\begin{align}
\phi(p_X,\tilde{p}_{\hat{X}^n})\leq P\quad \mbox{a.s.},
\end{align}
where $\tilde{p}_{\hat{X}^n}$ denotes the empirical distribution of $\hat{X}^n$.
For these two variants, however, the machinery of classical rate-distortion theory \cite[Theorem 9]{CYWSGT22} \cite{WO05}, suffices to establish \eqref{eq:zeroC}.


\subsection{Operational Implications}

The Blau--Michaeli function offers interesting operational guidelines for perceptual compression, particularly when $d(x,\hat{x})$ is the squared-error distortion measure $\|x-\hat{x}\|^2$ and $\phi(p_X,p_{\hat{X}})$ is the squared Wasserstein-$2$ distance
\begin{align}
W^2_2(p_X,p_{\hat{X}}):=\inf\limits_{\pi\in\Pi(p_X,p_{\hat{X}})}\mathbb{E}_{\pi}[\|X-\hat{X}\|^2],
\end{align}
where $\Pi(p_X,p_{\hat{X}})$ is the set of all  couplings of $p_X$ and $p_{\hat{X}}$.

Let $D_{\mathrm{G}}(R,P)$ denote the Blau--Michaeli  function for a Gaussian source $X\sim\mathcal{N}(\mu,\sigma^2)$ under squared-error distortion and squared Wasserstein-$2$ perception measures. As shown in \cite[Theorem 1]{ZQCK21},
\begin{align}
D_{\mathrm{G}}(R,P)=\sigma^22^{-2R}+\left[(\sigma(1-\sqrt{1-2^{-2R}})-\sqrt{P})_+\right]^2.\label{eq:scalarGaussian}
\end{align}
An important insight from this characterization is that, for a fixed rate, the reconstruction variables $\hat{X}$ corresponding to different points on the optimal distortion-perception tradeoff curve are related through a simple scaling operation. This suggests that, for Gaussian sources, the entire distortion-perception frontier can be achieved by fixing the encoder and varying only the decoder. This observation leads to the important notion of a universal representation, whose existence has been demonstrated, in an approximate sense, both analytically and empirically for general source distributions.

Another important finding \cite{Zou06} is that, for a general source with mean $\mu$ and variance $\sigma^2$, under squared-error distortion and squared Wasserstein-$2$ perception measures,
\begin{align}
D(R,P)\leq D_{\mathrm{G}}(R,P). \label{eq:worstcase}
\end{align}
This can be viewed as a generalization of the well-known worst-case property of Gaussian sources in the classical rate-distortion setting, which corresponds to $P=\infty$. In particular, \eqref{eq:worstcase} implies that, for a general source, the perceptual quality at a given rate can be adjusted simply by scaling the reconstruction corresponding to the perfect-perception point ($P=0$), while retaining the distortion guarantee associated with the corresponding Gaussian source.

Note that \eqref{eq:scalarGaussian} extends naturally to the vector Gaussian source case, yielding a generalization of the celebrated reverse water-filling formula. Interestingly, the resulting rate allocation exhibits a fundamental difference from the classical rate-distortion solution. In the absence of a perception constraint, a common water level determines the rate allocation, and signal dimensions with eigenvalues below the water level are left uncoded. In contrast, when the perception constraint is active, every signal dimension is assigned a strictly positive rate, regardless of its eigenvalue \cite{SSK24,QSCKYSGT25}. This phenomenon suggests that the perception constraint fundamentally changes the rate allocation strategy, forcing the encoder to preserve even low-energy components---often corresponding to high-frequency content---that would otherwise be discarded under conventional distortion-based compression.

In practice, the perception constraint is typically enforced using generative techniques---such as GANs, diffusion models, and flow matching---integrated into neural network-based compression pipelines. As the preceding discussion highlights, rate-distortion-perception theory provides not only theoretical performance benchmarks but also a crucial architectural blueprint, guiding the design of learned systems that operate near these fundamental limits.

\subsection{Computational Algorithms}

The computation of the Blau--Michaeli function $D(R,P)$ in \eqref{eq:objective}--\eqref{eq:single_perception} involves two limiting cases that connect to classical algorithms. In the absence of the perception constraint \eqref{eq:single_perception}, $D(R,P)$ reduces to Shannon's distortion-rate function $D(R)$, which can be computed using the Blahut--Arimoto algorithm. On the other hand, when the reconstruction distribution $p_{\hat X}$ is fixed, the optimization reduces to an entropic optimal transport problem, for which the Sinkhorn algorithm provides an efficient computational method. Thus, computing $D(R,P)$, in a certain sense, requires a synthesis of these two algorithmic frameworks.

In \cite{CNYWBCL23}, the computation of $D(R,P)$ is reformulated as a Wasserstein barycenter problem; an alternating Sinkhorn algorithm is then proposed to solve its entropy-regularized version, whose solution converges to that of the original Wasserstein barycenter problem as the regularization vanishes. 
In \cite{SSK25}, alternating minimization schemes are proposed for computing $D(R,P)$ by resolving the implicit equations induced by the perception constraint in a Blahut--Arimoto-type iteration. 
Despite these advances, much remains to be done to generalize the principles underlying both the Blahut--Arimoto and Sinkhorn algorithms and develop efficient, broadly applicable methods for computing the Blau-Michaeli function.










\section{Alternative Formulations}

\subsection{Sequence-Level Perception Constraints}

Setting $P=0$ in \eqref{eq:perception} ensures only symbol-level distributional consistency, namely, $p_{\hat{X}t}=p_X$ for $t=1,2,\ldots,n$. In some applications, however, it is preferable to impose the more stringent sequence-level distributional consistency condition $p_{\hat{X}^n}=p_{X^n}$. This motivates replacing \eqref{eq:perception} with
\begin{align}
\frac{1}{n}\phi(p_{X^n},p_{\hat{X}^n})\leq P.\label{eq:sequenceP}
\end{align}
However, a generic divergence $\phi$ is not amenable to information-theoretic analysis. Consequently, attention is often restricted to perception measures induced by optimal transport, namely,
\begin{align}
\phi(p_{X^n},p_{\hat{X}^n})=\inf\limits_{\pi\in\Pi(p_{X^n},p_{\hat{X}^n})}\sum\limits_{t=1}^nc(X_t,\hat{X}_t),\label{eq:ot}
\end{align}
where $c:\mathcal{X}\times\hat{\mathcal{X}}\rightarrow\mathbb{R}+$ is a cost function. Let $D_{\mathrm{J}}(R,P,C)$ denote the counterpart of $D_{\mathrm{M}}(R,P,C)$ obtained by replacing \eqref{eq:perception} with \eqref{eq:sequenceP}, where the subscript $\mathrm{J}$ indicates that perceptual quality is evaluated in terms of the joint distributions of the source and reconstruction sequences. Throughout this section, we assume that the perception measure takes the form given in \eqref{eq:ot}.

So far, a complete characterization of $D_{\mathrm{J}}(R,P,C)$ has been obtained only for the case of squared error distortion measure and squared Wasserstein-$2$ perception measure (corresponding to $c(x,\hat{x})=\|x-\hat{x}\|^2$). Specifically, we have \cite[Theorem 1]{QCYX25}
\begin{align}
D_{\mathrm{J}}(R,P,C)=&\inf\limits_{p_{\tilde{X}}}\inf\limits_{\pi_1,\pi_2\in\Pi(p_X,p_{\tilde{X}})}\mathbb{E}[\|X-\tilde{X}\|^2]\nonumber\\
&+\left[(\sqrt{\mathbb{E}_{\pi_1}[\|X-\tilde{X}\|^2]}-\sqrt{P})_+\right]^2\\
&\mbox{s.t.}\quad\mathbb{E}_{\pi_1}[X|\tilde{X}]=\tilde{X}\quad\pi_1-\mbox{a.s.},\\
&\hspace{0.3in}I_{\pi_1}(X;\tilde{X})\leq R,\\
&\hspace{0.3in}I_{\pi_2}(X;\tilde{X})\leq R+C.
\end{align}
The interpretation of the above characterization is somewhat involved; see \cite{HCTP26} for a detailed exposition. Here, we focus on the special case of no common randomness, for which $D_{\mathrm{J}}(R,P,C)$ simplifies to
\begin{align}
D_{\mathrm{J}}(R,P,0)=D(R)+\left[(\sqrt{D(R)}-\sqrt{P})_+\right]^2.\label{eq:C=0}
\end{align}
Note that setting $P=\infty$ in \eqref{eq:C=0} recovers Shannon's distortion-rate function:
\begin{align}
D_{\mathrm{J}}(R,\infty,0)=D(R).\label{eq:DR}
\end{align}
On the other hand, setting $P=0$ in \eqref{eq:C=0} gives
\begin{align}
D_{\mathrm{J}}(R,0,0)=2D(R),\label{eq:doubling}
\end{align}
which is exactly twice the minimum achievable distortion in the absence of a perception constraint. In particular, \eqref{eq:doubling} can be achieved by using a conventional distortion-based encoder attaining \eqref{eq:DR} and setting the decoder to its stochastic inverse, i.e., performing posterior sampling \cite{YWYML21,LZCK22,LZCK22J}. Linearly interpolating between the two extreme reconstructions corresponding to $P=\infty$ and $P=0$ then yields the optimal distortion-perception tradeoff in \eqref{eq:C=0} \cite{FMM21,YWL22}.


\subsection{IID Reconstruction Constraints}

One can make the problem more tractable by additionally requiring the reconstruction sequence to be i.i.d., thereby connecting perceptual compression to the output-constrained lossy source coding problem studied in \cite{SLY15J2}. Let $D_{\mathrm{I}}(R,P,C)$ be defined analogously to $D_{\mathrm{J}}(R,P,C)$ with an extra i.i.d. constraint on the  reconstruction sequence, where the subscript $\mathrm{I}$ indicates the independence requirement. Note that if the reconstruciton sequence $\hat{X}^n$ is i.i.d. with marginal distribution $p_{\hat{X}}$, then  
$\frac{1}{n}\phi(p_{X^n},p_{\hat{X}^n})=\phi(p_X,p_{\hat{X}})$.
Consequently, it follows from \cite[Theorem 1]{SLY15J2} that
\begin{align}
D_{\mathrm{I}}(R,P,C)=&\inf\limits_{p_{U\hat{X}|X}}\mathbb{E}[d(X,\hat{X})]\\
&\mbox{s.t.}\quad X\leftrightarrow U\leftrightarrow\hat{X}\mbox{ is Markovian},\\
&\hspace{0.3in}I(X;U)\leq R,\\
&\hspace{0.3in}I(\hat{X};U)\leq R+C,\\
&\hspace{0.3in}\phi(p_X,p_{\hat{X}})\leq P.
\end{align}
Clearly, we have
\begin{align}
D_{\mathrm{I}}(R,P,C)\geq D_{\mathrm{J}}(R,P,C)\geq D_{\mathrm{M}}(R,P,C).
\end{align}
Moreover, with unlimited common randomness, all three formulations coincide with the Blau--Michaeli  function $D(R,P)$. See \cite{XLCZ25,XLCYZ26} for bounds on $D_{\mathrm{I}}(R,P,C)$ in the Gaussian setting with squared-error distortion and squared Wasserstein-$2$ perception measures.


\subsection{Conditional Distributional Perception Constraints}

Let $D_{\mathrm{Cn}}(R,P,C)$ be defined by replacing \eqref{eq:sequenceP} with
\begin{align}
\frac{1}{n}\phi(p_{X^n|J},p_{\hat{X}^n|J})\leq P,
\end{align}
where the subscript $\mathrm{Cn}$ indicates that perceptual quality is evaluated in terms of the conditional distributions of the source and reconstruction sequences given
the encoder output.
A computable characterization of $D_{\mathrm{Cn}}(R,P,C)$ is known only for the case without common randomness \cite[Theorem 1]{SCKY24}:
\begin{align}
D_{\mathrm{Cn}}(R,P,0)=&\inf\limits_{p_{U\hat{X}|X}}\mathbb{E}[d(X,\hat{X})]\\
&\mbox{s.t.}\quad X\leftrightarrow U\leftrightarrow\hat{X} \mbox{ is Markovian},\\
&\hspace{0.3in}I(X;U)\leq R,\\
&\hspace{0.3in}\mathbb{E}_{U}[\phi(p_{X|U},p_{\hat{X}|U})]\leq P.
\end{align}
Since 
$\phi(p_{X^n|J},p_{\hat{X}^n|J})\geq \phi(p_{X^n},p_{\hat{X}^n})$, it follows that
\begin{align}
D_{\mathrm{Cn}}(R,P,C)\geq D_{\mathrm{J}}(R,P,C),\label{eq:cvsj}
\end{align}
where equality holds in \eqref{eq:cvsj} under squared-error distortion and squared Wasserstein-$2$ perception measures when $C=0$.



\section{Extension to Video Coding}

Extending the rate-distortion-perception framework to sequential video compression poses additional challenges in the choice of perception loss functions. For example, one can enforce distributional consistency between the source and reconstruction either with respect to the joint distribution of all video frames up to the current frame or with respect to the framewise marginal distributions. It is demonstrated in \cite{SPCYK23} that the former choice provides better temporal consistency across video frames but suffers from the phenomenon of error permanence, whereby reconstruction errors propagate to future frames;  in contrast, the latter choice offers greater flexibility in correcting reconstruction errors over time.

A self-adaptive perception loss function is proposed in \cite{SPCDYCK26} that enforces consistency between the current source frame and its reconstruction through their respective joint distributions with the preceding reconstructions. This approach can simultaneously mitigate the phenomenon of error permanence and better exploit the temporal correlations among high-quality reconstructions.





\vfill\pagebreak

\bibliographystyle{IEEEbib}

\begin{thebibliography}{1}













\bibitem{BM19} Y. Blau and T. Michaeli, ``Rethinking lossy compression:
The rate-distortion-perception tradeoff," in {\em Proc. ACM Int. Conf. Mach. Learn. (ICML)}, 2019, pp. 675--685.

\bibitem{TW21} L. Theis and A. B. Wagner, ``A coding theorem for
the rate-distortion-perception function," in {\em Proc. Neural Compress. Workshop Int. Conf. Learn. Represent. (ICLR)}, 2021, pp. 1--5.


\bibitem{CYWSGT22}	
J. Chen, L. Yu, J. Wang, W. Shi, Y. Ge, and W. Tong, ``On the rate-distortion-perception function," {\em IEEE J. Sel. Areas Inf. Theory}, vol. 3, no. 4, pp. 664--673, Dec. 2022.

\bibitem{LEG18} C. T. Li and A. El Gamal, ``Strong functional representation
lemma and applications to coding theorems," {\em IEEE Trans. Inf.
		Theory}, vol.~64, no. 11, pp.~6967--6978, Nov. 2018.

        \bibitem{WO05}
T. Weissman and E. Ordentlich, ``The empirical distribution of rate-constrained source codes," {\em IEEE Trans. Inf. Theory}, vol.~51, no.~11, pp.~3718--3733, Nov. 2005.


\bibitem{ZQCK21} G. Zhang, J. Qian, J. Chen, and A. Khisti, "Universal
rate-distortion-perception representations for lossy compression,"
in {\em Proc. Adv. Neural Inf. Process. Syst. (NeurIPS)}, 2021, pp.~11517--11529.


\bibitem{Zou06}
Zou et al., ``Gaussian extremality from self-couplings," preprint.

\bibitem{SSK24}
G. Serra, P. A. Stavrou, and M. Kountouris, ``On the computation of the Gaussian rate–distortion–perception function," {\em IEEE J. Sel. Areas Inf. Theory}, vol.~5, pp. 314--330, 2024.

\bibitem{QSCKYSGT25}
J. Qian, S. Salehkalaibar, J. Chen, A. Khisti, W. Yu, W. Shi, Y. Ge, and W. Tong, ``Rate-distortion-perception tradeoff for vector Gaussian sources," {\em IEEE J. Sel. Areas Inf. Theory}, vol. 6, pp. 1--17, 2025.



\bibitem{CNYWBCL23}
C. Chen, X. Niu, W. Ye, S. Wu, B. Bai, W. Chen, and S. Lin, ``Computation of rate-distortion-perception functions with Wasserstein barycenter,"
in {\em Proc. IEEE Int. Symp. Inf. Theory (ISIT)}, 
2023, pp. 1074--1079.


\bibitem{SSK25}
G. Serra, P. A. Stavrou, and M. Kountouris, ``Alternating minimization schemes for computing rate-distortion-perception functions with $f$-divergence perception constraints," {\em IEEE Trans. Inf. Theory}, vol.~71, no.~11, pp.~9100--9115, Nov. 2025.



\bibitem{QCYX25}
X. Qu, J. Chen, L. Yu and X. Xu, ``Rate-distortion-perception theory for the quadratic Wasserstein space," {\em IEEE Trans. Inf. Theory}, vol.~71, no.~11, pp.~8247--8261, Nov. 2025.

\bibitem{HCTP26}
A. Hussein, J. Chen, C. Tian, and S. S. Pradhan, ``Constructive approaches to perception-aware lossy source soding: Information-theoretic guidelines,"  arXiv:2604.19515.

\bibitem{YWYML21} Z. Yan, F. Wen, R. Ying, C. Ma, and P. Liu, ``On perceptual lossy compression: The cost of perceptual reconstruction and an optimal training framework," in {\em Proc. ACM Int. Conf. Mach. Learn. (ICML)}, 2021, pp. 11682--11692.


\bibitem{LZCK22} H. Liu, G. Zhang, J. Chen, and A. Khisti, ``Lossy compression
with distribution shift as entropy constrained optimal transport,"
in {\em Proc. Int. Conf. Learn. Represent. (ICLR)},
2022, pp.~1--34.



\bibitem{LZCK22J}
H. Liu, G. Zhang, J. Chen, and A. Khisti, ``Cross-domain lossy compression as entropy constrained optimal transport," {\em IEEE J. Sel. Areas Inf. Theory}, vol.~3, no.~3, pp. 513--527, Sep. 2022.

\bibitem{FMM21}
D. Freirich, T. Michaeli, and R. Meir, ``A theory of the distortion-perception tradeoff in Wasserstein space,” in {\em Proc. Adv.
Neural Inf. Process. Syst. (NeurIPS)}," 2021, pp. 25661--25672.



\bibitem{YWL22}
Z. Yan, F. Wen, and P. Liu, ``Optimally controllable perceptual lossy compression," in {\em Proc. ACM Int. Conf. Mach. Learn.
(ICML)}, 2022, pp. 24911--24928.


\bibitem{SLY15J2} N. Saldi, T. Linder, and S. Y\"{u}ksel, ``Output constrained
lossy source coding with limited common randomness," {\em IEEE
	Trans. Inf. Theory}, vol.~61, no.~9, pp.~4984--4998, Sep. 2015.


\bibitem{XLCZ25}
L. Xie, L. Li, J. Chen, and Z. Zhang, ``Output-constrained lossy source coding with application to rate-distortion-perception theory," {\em IEEE Trans. Commun.},  vol. 73, no. 3, pp. 1801--1815, Mar. 2025.


\bibitem{XLCYZ26}
L. Xie, L. Li, J. Chen, L. Yu, and Z. Zhang, ``Gaussian rate-distortion-perception coding and entropy-constrained scalar quantization,"  {\em IEEE Trans. Commun.}, vol. 74, pp. 3298--3312, 2026.

\bibitem{SCKY24}
S. Salehkalaibar, J. Chen, A. Khisti, and W. Yu, ``Rate-distortion-perception tradeoff based on the conditional-distribution perception measure," {\em IEEE Trans. Inf. Theory}, vol.~70, no.~12, pp.~8432--8454, Dec. 2024.






























\bibitem{SPCYK23}
S. Salehkalaibar, B. Phan,  J. Chen, W. Yu, and A. Khisti, ``On the choice of perception loss function for learned video compression," in {\em Proc. Adv. Neural Inf. Process. Syst. (NeurIPS)}, 2023, pp.~1--19.


\bibitem{SPCDYCK26}
S. Salehkalaibar, B. Phan, L. Cai, J. A. Dick, W. Yu, J. Chen, and A. Khisti, ``On self-adaptive perception loss function for sequential lossy compression," 	arXiv:2502.10628.



























































\end{thebibliography}

\end{document}